\documentclass[aip,jcp,reprint,showkeys,showpacs,longbibliography,noeprint]{revtex4-1} 
\usepackage[utf8]{inputenc}

\makeatletter
\preprintsty@sw{
  \usepackage{geometry}
  \geometry{inner=1.5cm, outer=4.75cm, marginparwidth=4cm}

  \usepackage[notcite,notref]{showkeys}
  
}{}
\makeatother

\usepackage{cmap} 
\usepackage[T1]{fontenc} 

\usepackage{lmodern}
\usepackage{microtype}

\usepackage{amsmath,amssymb,mathtools}
\usepackage{mleftright} 
\DeclareMathOperator{\cotanh}{coth}

\usepackage{graphicx,grffile}
\graphicspath{{figures/}}
\newlength{\figwidth}
\usepackage{url,hyperref}
\usepackage[natural,table,usenames,dvipsnames]{xcolor}
\hypersetup{colorlinks=true, linkcolor=BrickRed,
  urlcolor=blue!50!black, citecolor=blue!50!black}

\usepackage[capitalize]{cleveref}
\crefrangelabelformat{equation}{(#3#1#4)$-$(#5#2#6)}

\usepackage[version=4,font=]{mhchem}  

\usepackage{booktabs}

\usepackage[inline,shortlabels]{enumitem}
\setlist[enumerate]{label={(\roman*)},font=\itshape, itemjoin=\hspace{1ex}}

\newcommand{\kB}{{k_\text{B}}}

\renewcommand\leq{\leqslant}
\renewcommand\geq{\geqslant}
\renewcommand\rho\varrho
\renewcommand\epsilon\varepsilon
\renewcommand\phi\varphi
\renewcommand\vec[1]{\textrm{\bfseries #1}}

\let\d\diff
\newcommand\e{\mathrm{e}}

\begin{document}
\title{Diffusion-influenced reaction rates in the presence of pair interactions}

\author{Manuel Dibak}
\thanks{Equal contributions}
\affiliation{
    Freie Universität Berlin, Fachbereich Mathematik und Informatik,
    Arnimallee 6, 14195 Berlin, Germany
}

\author{Christoph Fröhner}
\thanks{Equal contributions}
\affiliation{
    Freie Universität Berlin, Fachbereich Mathematik und Informatik,
    Arnimallee 6, 14195 Berlin, Germany
}

\author{Frank Noé}
\email{frank.noe@fu-berlin.de}
\affiliation{
    Freie Universität Berlin, Fachbereich Mathematik und Informatik,
    Arnimallee 6, 14195 Berlin, Germany
}

\author{Felix Höf{}ling}
\email{f.hoefling@fu-berlin.de}
\affiliation{
    Freie Universität Berlin, Fachbereich Mathematik und Informatik,
    Arnimallee 6, 14195 Berlin, Germany
}
\affiliation{Zuse Institute Berlin, Takustr. 7, 14195 Berlin, Germany}

\date{\today}

\begin{abstract}
  The kinetics of bimolecular reactions in solution depends, among other
  factors, on intermolecular forces such as steric repulsion or electrostatic
  interaction. Microscopically, a pair of molecules first has to
  meet by diffusion before the reaction can take place.  In this work, we
  establish an extension of Doi's volume reaction model to
  molecules interacting via pair potentials,
  which is a key ingredient for
  interacting-particle-based reaction--diffusion (iPRD) simulations.
  As a central result, we relate model parameters and macroscopic reaction rate constants in this situation.
  We solve the corresponding reaction--diffusion equation in the steady state and derive
  semi-analytical expressions for the reaction rate constant and the
  local concentration profiles.
  Our results apply to the full spectrum from well-mixed to diffusion-limited kinetics.
  For limiting cases, we give explicit formulas, and we provide a
  computationally inexpensive numerical scheme for the general case, including
  the intermediate, diffusion-influenced regime.
  The obtained rate constants decompose uniquely into encounter and formation rates,
  and we discuss the effect of the potential on both subprocesses, exemplified for a soft harmonic repulsion and a Lennard-Jones potential.
  The analysis is complemented by extensive stochastic iPRD simulations, and we find excellent agreement with the theoretical predictions.
\end{abstract}

\maketitle

\newcommand\iPRD{iPRD}

\section{Introduction}

A microscopic view on bimolecular chemical reactions in solution is essential for our understanding of many biological processes
and technological applications; recent examples include, most prominently, protein functioning via complex formation \cite{Scott2016,Plattner2017}, ligand binding \cite{Houslay2010,Paul2017}, and oligomerisation \cite{Burre2014,Schoneberg2017},
and on the other hand, catalysis in nanoreactors \cite{Herves:2012,Galanti2016}
or ion deposition in batteries \cite{Zhou:AEM2017,Armand2008}.
Such reactions are often strongly influenced by diffusion of at least one reactant, even more if transport
occurs in a heterogeneous environment such as the interior of cells or on cellular membranes \cite{Melo:2006,Zhou:2008,Hoefling:2013,Weiss:2014}.

In eukaryotes, the intracellular space is densely crowded by macromolecules, meandered by filamental networks, and compartmentalized by extended organelles, typically rendering diffusion at small scales anomalous \cite{Etoc2018nonspecific, Witzel2019heterogeneities, Banks:2016,Stiehl2016heterogeneity, Kusumi:2005, Metzler:2016, AlbrechtEtAl_JCP16_Nanoscopic, Horton2010development}.
Different modelling strategies have been advised to account for such situations \cite{Smith:2018}:
spatio-temporal master equations exploit metastability of diffusion between compartments \cite{Winkelmann:2016},
and crowding has been incorporated into the reaction--diffusion master equation on a mesoscale level \cite{Engblom:2018}.
In particle-based Brownian dynamics simulations, crowding is implemented frequently as explicit excluded volume
via hard or short-range repulsions \cite{Ridgway:2008,Kim:2010,Dorsaz:2010,Grima:2010,Trovato:2014,Echeverria:2015},
which can give rise to complex-shaped structures on a cascade of scales \cite{Lorentz_JCP:2008,Spanner:2016,Schnyder2015rounding,Petersen2019anomalous}.

Stochastic particle–based reaction diffusion simulations have become increasingly popular in the past decade~\cite{Morelli2008,Erban2009,Johnson2014,Schoneberg2014,schoneberg:2014simulation,Vijaykumar2015,Andrews2016,Michalski:2016springsalad,Arjunan2017multi,Sadeghi2018,andrews2018particle}. Such simulation methods and frameworks evolve the reaction–diffusion processes microscopically and have experienced advancements both in accuracy and computational performance~\cite{Donev2018,frohner2018reversible,dibak2018msm,sbailo2017efficient,Sbailo2019formalization}.
A recent development is interacting particle reaction dynamics (iPRD)~\cite{Schoneberg2013,biedermann2015readdymm,Hoffmann2018} that allows general interaction potentials on the reactive particles, for example, steric reuplsion or electrostatic forces.
Such interaction potentials may represent free energy landscapes computed from molecular dynamics (MD) simulations~\cite{buch2011optimized,Xu:2019,wu2016multiensemble}.

A bimolecular reaction, \ce{A + B -> X}, of two molecules \ce{A} and \ce{B} in solution occurs as a two-step process: encounter of the two reacting molecules by diffusion, followed by the formation of the product \ce{X}, which abbreviates, for example, a complex \ce{C} or the result \ce{A^* + B} of a catalytic reaction.
Statistical independence of the durations of both steps suggests that the total reaction rate constant $k$ is the harmonic mean~\cite{Shoup1982,Szabo1980} of an encounter rate $k_\mathrm{e}$ and a formation rate $k_\mathrm{f}$:
\begin{equation}
    \label{eq:total-rate-constant}
    k^{-1} = k_\mathrm{e}^{-1} + k_\mathrm{f}^{-1}.
\end{equation}
The formation rate depends on the detailed chemistry of the reaction process, often pictured as surmounting an activation barrier, whereas the encounter rate is determined by spatial diffusion of the molecules and subject to crowding conditions \cite{Kim:2010,Dorsaz:2010,Grima:2010,Echeverria:2015}, interaction potentials \cite{Debye1942c}, and confining geometries \cite{Grebenkov2018strong}.
A \emph{diffusion-influenced} reaction refers to the not uncommon situation that both rates in \cref{eq:total-rate-constant} are of comparable magnitude and both steps are relevant for the overall kinetics~\cite{Bhalla2004}.

A commonly used reaction scheme in \iPRD\ is Doi's volume reaction
model~\cite{Teramoto1967,Doi1975,Doi1975b,Doi1976stochastic}, where a reaction can occur with
a microscopic rate $\lambda$ if molecule centres are within a reaction radius $R$.
Here, we extend this scheme by a pair interaction and relate the model parameters $\lambda$ and $R$ to the macroscopic reaction rate and its components for encounter and formation, see \cref{eq:total-rate-constant}.
Inversion of such a relation would allow the calibration of the microscopic model to match
experimental rates.
We obtain insights into the specific contributions of attractive and repulsive
interactions to the reaction kinetics, and we highlight the importance of the local concentration
of molecules in the reaction zone, which may differ drastically from the equilibrium distribution.

\section{Microscopic model}
\label{sec:model}

Microscopic theories for bimolecular reactions date back to Smoluchowski~\cite{Smoluchowski1917} in 1917, who proposed and analysed a model for coagulation of sphere-like molecules in solution that react instantaneously upon contact.
Later, Debye~\cite{Debye1942c} amended the model by electrostatic interactions between the reactants, with notable repercussions on the binding rate.
Collins and Kimball~\cite{Collins1949a,Collins1949b} refined Smoluchowski's model by introducing a finite rate at which molecules would react on contact.
This model has been widely studied in the literature~\cite{Shoup1982,Szabo1980,Agmon1990,DelRazo2018},
however, the singular nature of the reaction surface has drawbacks in computer simulations as the exact time of encounter is not resolved in a time-stepping algorithm.
An alternative scheme was suggested by Teramoto and Shigesada~\cite{Teramoto1967} and further characterized by Doi~\cite{Doi1975,Doi1975b,Doi1976stochastic},
which permits the reaction of two molecules with a microscopic rate $\lambda$, referred to as \emph{propensity} \cite{Gillespie2007stochastic}, as long as the reactants are within a reaction radius $R$.
This model is often referred to as the \textit{volume reaction model} or \textit{Doi model}
and is in the focus of the present study.

\begin{figure}
    \centering
    \includegraphics[width=\figwidth]{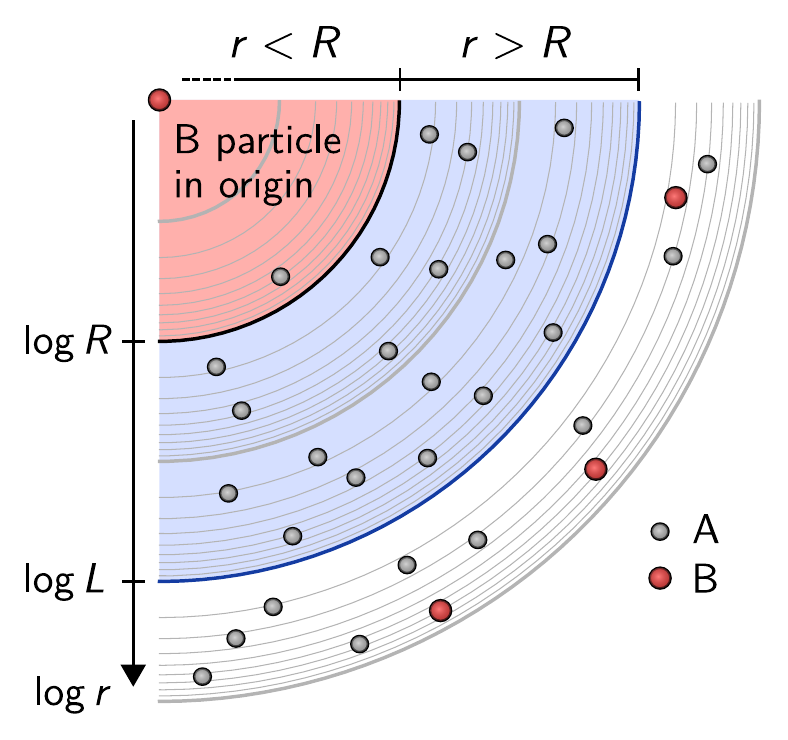}
    \caption{System of reactive molecules. Molecules of species \ce{A} diffuse in space and can react with \ce{B} molecules if their distance $r$ is smaller than the reaction radius $R$. If \ce{B} particles are scarce, a reasonable assumption is that there is no competition between them and one can treat only one of them within a spherical domain of radius $L \gg R$.
    For the analytical treatment, $L\to\infty$, whereas for numerical methods and simulations $L$ is finite.
    }
    \label{fig:scheme}
\end{figure}

Following Smoluchowski~\cite{Smoluchowski1917}, we consider a solution of substances \ce{A} and \ce{B}, that undergo the reaction
\begin{equation}
    \label{eq:the-reaction}
    \ce{A + B -> A^* + B},
\end{equation}
for which the product \ce{A^*} of the reaction falls out of scope, such that we do not need to consider it.
The concentrations $c_A$ and $c_B$ of \ce{A} and \ce{B} molecules, respectively, are assumed to be both so dilute that interactions between like molecules can safely be ignored.
(Otherwise, the reaction kinetics would non-trivially depend on $c_A$ and $c_B$ and the reaction rate would not be a well-defined constant.)
Further, the concentration of \ce{B} molecules is assumed to be much smaller than that of \ce{A}, $c_B \ll c_A$,
i.e., \ce{A} molecules are abundant relative to \ce{B}s and there is no competition for reactants between the \ce{B} molecules.
Equivalently, substance \ce{B} is highly diluted, and the problem can be rephrased as that of a single \ce{B} molecule surrounded by \ce{A} molecules in a large, yet finite volume $V$.
It is convenient to switch to the reference frame of the \ce{B} molecule, and we will choose a spherical volume $V$ of radius~$L$; see \cref{fig:scheme} for an illustration.
In a finite amount of time and for sufficiently large $V$, the \ce{B} molecule absorbs only a negligible fraction of \ce{A}s
so that we can assume a quasi-steady state with the concentration $c_A$ being constant at the boundary $\partial V$ of the volume.

As microscopic reaction model, we use the Teramoto--Shigesada--Doi model \cite{Teramoto1967,Doi1975,Doi1975b,Doi1976stochastic}, in which \ce{A} and \ce{B} molecules diffuse in space with diffusion constants $D_A$ and $D_B$, respectively, forming a reactive complex whenever an \ce{A} is separated from a \ce{B} by less than the reaction distance $R$.
This reactive complex undergoes reaction \eqref{eq:the-reaction} with a microscopic rate constant or \emph{propensity} $\lambda$, thus effectively removing \ce{A} molecules from the system with a frequency $K$.
More precisely, given a reactive complex, reaction events are triggered by a Poisson clock with parameter $\lambda$.
The throughput or velocity of reaction \eqref{eq:the-reaction} is then given by
\begin{equation}
    \label{eq:throughput-micro}
    \frac{\mathrm{d}c_{A^*}}{\mathrm{d}t} = K c_{B} \,,
\end{equation}
where $c_{A^*}$ is the overall concentration of the reaction product \ce{A^*}.

Similarly to Debye's work~\cite{Debye1942c}, and as commonly done in \iPRD\ simulations~\cite{Schoneberg2013}, our focus here is on situations where \ce{A} and \ce{B} molecules interact physically with each other according to an isotropic pair potential $U(\vec r) = U(|\vec r|)$; the vector $\vec r$ denotes the separation of an \ce{AB} pair.
The average concentration field $p(\vec r,t)$ of \ce{A} molecules and the corresponding flux (density) $\vec j(\vec r, t)$ are then governed by the reaction--diffusion equation
\begin{subequations}
    \label{eq:reaction-diffusion}
\begin{align}
    \label{eq:reaction-diffusion-balance}
    \partial_{t} p(\vec r, t) &= - \nabla \cdot  \vec j(\vec r, t) - a(\vec r) \, p(\vec r, t) \,,\\
    \label{eq:flux}
    \vec j(\vec r,t) &:= -D \, \e^{-\beta U(\vec r)}\nabla \mleft[ \e^{\beta U(\vec r)} p(\vec r, t)\mright] \,,
\end{align}
\end{subequations}
with the reaction propensity $a(\vec r) \geq 0$ and $D=D_A+D_B$ the relative diffusion constant of the particles;
$\beta = 1/\kB T$ denotes the inverse of the thermal energy scale as usual.
Within the Doi model, the propensity $a(\vec r)$ is implemented in terms of the Heaviside step function,
$a(\vec r) = \lambda \,\theta(R-|\vec r|)$ such that the \ce{B} molecule appears as a spherical reactive sink of radius~$R$.

By isotropy of the setup, the steady flux $\vec{j}(\vec r)$ of \ce{A} molecules has only a radial component $j(r)$ that is a function only of the distance $r=|\vec r|$ to the \ce{B} molecule.
It determines the reaction frequency $K$ through the surface integral
\begin{equation}
    \label{eq:production-rate}
    K = -\int _{|\vec r| = R} \vec j(\vec r) \cdot \vec n \,\d\sigma = -4 \pi R^{2} j(R) \,,
\end{equation}
with the surface normal $\vec n$ pointing outwards;
the minus sign arises due to the fact that particles flow from the boundary to the sink at the origin, $j(r)<0$.
On the other hand, the law of mass action yields the reaction rate equation
\begin{align}
    \label{eq:throughput-macro}
    \frac{\d c_{A^*}}{\d t} = k c_{A}c_{B} \,,
\end{align}
in terms of the macroscopic association rate constant~$k$.
Comparing to \cref{eq:throughput-micro}, the latter is related to the microscopic frequency $K$ by
$k = K / c_A$, and the reaction rate constant follows as
\begin{equation}
    \label{eq:relation-k-and-flux}
    k = \frac{4 \pi R^{2} |j(R)|}{c_A} \,.
\end{equation}

The goal of the following sections is to calculate the flux profile $j(r)$ of the quasi-steady state and thus the macroscopic rate $k$, focussing on their dependences on the microscopic reaction parameters, $\lambda$ and $R$, and on the pair potential $U(r)$ between \ce{A} and \ce{B} molecules. Note that there is no interaction amongst \ce{A} molecules.

\section{Solution strategy and classical limiting cases}
\label{sec:solution-strategy}

In this section, we work out the general solution strategy for the reaction--diffusion equations, \cref{eq:reaction-diffusion}, and obtain analytical solution to important subproblems, which resemble a number of classical results.
The stationary solutions $p(\vec r)$ obeys $\partial_t p(\vec r) = 0$, and thus \cref{eq:reaction-diffusion-balance} reduces to
\begin{equation}
    \label{eq:rd-equation}
    \nabla \cdot \vec j(\vec r) = -a(\vec r) \, p(\vec r).
\end{equation}
According to the quasi-steady state assumption, $p(\vec r)$ further satisfies the Dirichlet boundary condition
\begin{equation}
    \label{eq:upper-bc-concentration}
    p(\vec r) = c_A \,, \quad \vec r \in \partial V \,.
\end{equation}
Restricting to isotropic potentials, we switch to a single radial coordinate, $r = |\vec r|$, with the convention that
the flux $j(r) = \vec j(\vec r) \cdot \vec r / r$ points outwards:
\begin{align}
    \label{eq:radialSmoluchowski}
    \frac{1}{r^{2}} \,\partial_{r} r^{2} j(r) &= -\lambda \, \theta(R-r) \, p(r)  \\
\intertext{with}
    \label{eq:radialCurrent}
    j(r) &= -D \e^{-\beta U(r)} \partial_{r} \mleft[ \e^{\beta U(r)} p(r) \mright] .
\end{align}
In this case and for an infinitely large volume~$V$, \cref{eq:upper-bc-concentration} simplifies to $p(r\to \infty) = c_A$.

To complete the boundary value problem for $p(r)$, we need to specify also the behaviour at the coordinate origin, which is not obvious due to the interaction potential.
The total flux through a ball $B_\epsilon$ of radius $\epsilon$ centred at $\vec r=0$ obeys:
\begin{equation}
  \int_{\partial B_\epsilon} \vec j(\vec r) \cdot \vec n\, \d\sigma
  = - \int_{B_\epsilon} a(\vec r) \, p(\vec r) \, \d^3 r \,,
\end{equation}
invoking Gauss' theorem and inserting \cref{eq:rd-equation}.
Continuity of the solution $p(\vec r)$ together with our choice for $a(\vec r)$ yields
$4\pi \epsilon^2 \, j(\epsilon) \simeq - \lambda p(0) \cdot 4\pi \epsilon^3/3$,
and thus
\begin{equation}
  \label{eq:lower-bc-flux}
  j(0) = 0 \,.
\end{equation}
It implies a Robin boundary condition for the concentration profile,
\begin{equation}
  \lim_{r \to 0} \bigl[\beta U'(r) \, p(r) + \partial_r p(r) \bigr] = 0,
\end{equation}
which is satisfied by a Boltzmann distribution (scaled by a constant factor):
\begin{equation}
  \label{eq:lower-bc-conc}
  p(r) \sim \exp\mathopen{}\mathclose{\boldsymbol({-\beta U(r)}\boldsymbol)} \,,  \quad r \to 0,
\end{equation}
capturing the $r$-dependence asymptotically.

Note that the preceding derivation does not apply for potentials $U(r)$ that diverge as $r \to 0$.
In this case, the current $\vec j(\vec r)$ is not defined at the origin, $\vec r=0$, and, strictly speaking, this point must be excluded from the integration domain $B_\epsilon$, which forbids the application of Gauss' theorem.
Yet, the extension of \cref{eq:lower-bc-conc} to diverging potentials, $U(r \to 0) = +\infty$, is motivated physically as it is improbable that any $A$ molecule reaches the centre of the reaction volume:
an upper bound on $p(r)$ is given by the equilibrium distribution, describing the non-reacting case.
In particular, $p(\vec r)$ is continuous in $\vec r = 0$ and so is $\nabla \cdot \vec j(r)$  by \cref{eq:rd-equation}, justifying the use of Gauss' theorem \emph{a posteriori}.

Eventually, the step-like reaction propensity in \cref{eq:radialSmoluchowski} suggests to split the domain at the reaction boundary, $r=R$, and to find separate solutions $p_\gtrless$ and $j_\gtrless$ in both subdomains, $r \gtrless R$.
By inspection of the r.h.s. of \cref{eq:radialSmoluchowski,eq:radialCurrent}, the flux $j(r)$ is finite and continuous at this interface, which implies that $p(r)$ is continuously differentiable at $r=R$. This provides us with the interface conditions
\begin{gather}
    \label{eq:cont-bc-conc}
    p_>(R) = p_<(R) \,,  \\
    \label{eq:cont-bc-flux}
    j_>(R) = j_<(R) = -K/4\pi R^2 ,
\end{gather}
making use of \cref{eq:production-rate} in the last step. Matching the solutions of both subdomains will thus yield the sought-after reaction frequency~$K$.

\subsection{Outer solution}

In the outer domain ($>$), where $R \leq r < \infty$, \cref{eq:radialSmoluchowski} reduces to an equation for the flux alone,
$\partial_{r} r^{2} j_>(r) = 0$. Integration from the lower boundary, \cref{eq:cont-bc-flux}, to some $r > R$ yields:
\begin{equation}
    \label{eq:outerFlux}
    j_>(r) = -\frac{K}{4\pi r^{2}} \,,
\end{equation}
with unknown rate $K$.
The functional dependence on $r$ is readily understood by the fact that, in the absence of reactions, the integral flux through spheres of radius $r$ is constant (Gauss' theorem).
In particular, the solution is compatible with the no-flux condition, $j_>(r \to \infty) = 0$, which is implied by the upper boundary, $p_>(r\to \infty) = c_A$, together with the vanishing force, $-\nabla U(r\to \infty) = 0$, and
using \cref{eq:radialCurrent}.

Next, we calculate the concentration profile $p_>(r)$ from \cref{eq:radialCurrent,eq:upper-bc-concentration}.
Introducing
\begin{equation}
  \label{eq:def-gr}
  g(r):=\e^{\beta U(r)} r^{-2}
\end{equation}
for brevity, one finds
$(K/4\pi D) \, g(r) = \partial_r \mleft[ \e^{\beta U(r)} p_>(r)\mright ]$,
and after integration over $[r,\infty)$:
\begin{equation}
  \label{eq:outerSolution}
  p_>(r) = \e^{-\beta U(r)}\left[ c_{A} - \frac{K}{4\pi D} \int _{r} ^{\infty} g(s) \d s \right ] \,,
\end{equation}
which is Debye's classical result \cite{Debye1942c}.
If the interaction potential is not present ($U=0$), this reduces to the familiar solution of the Dirichlet--Laplace problem:
\begin{equation}
  \label{eq:outerSolutionFree}
  p_>(r) = c_{A} - \frac{K}{4\pi D} \frac{1}{r} \,.
\end{equation}

For diffusion-limited reactions, that is when product formation is fast and $k_\mathrm{f} \ll k_\mathrm{e}$ in \cref{eq:total-rate-constant}, particles almost surely react on the surface of the reaction volume
and the concentration inside vanishes: $p_<(r)=0$ for $r \leq R$.
Then by continuity of $p(r)$ at the interface of the subdomains, \cref{eq:outerSolution} is amended by $p_>(R)=0$ and can be solved for $K$.
This yields the Debye reaction rate constant $k=K/c_A$, which we identify as the encounter rate $k_\mathrm{e}$ in the presence of a pair potential:
\begin{equation}
  k_\mathrm{e} = 4 \pi D \Big / \int _{R} ^{\infty} g(s) \d s \,.
  \label{eq:DebyeRate}
\end{equation}
The corresponding concentration profile is given by \cref{eq:outerSolution} and reads
\begin{equation}
  \label{eq:outer-solution-debye}
  p_>(r) = c_{A} \e^{-\beta U(r)} \int _{R} ^{r} g(s) \, \d s \Big / \int _{R} ^{\infty} g(s) \, \d s .
\end{equation}
In particular, $p_>(r)$ is independent of the diffusion constant~$D$.
For $U(r) = 0$, these results recover Smoluchowski's rate constant \cite{Smoluchowski1917} $k=4\pi DR$ and the profile
$p_>(r) = c_{A} (1 -R/r)$.

\subsection{Inner solution without potential}
\label{sec:innerSolutionWithout}

In the absence of an interaction potential, \cref{eq:radialSmoluchowski,eq:radialCurrent} simplify drastically and
the concentration inside $p_<(r)$ the reaction volume, $0 \leq r \leq R$, obeys the Helmholtz equation
\begin{equation}
    \label{eq:reaction-diffusion-inner}
    \mleft(\partial_r^2  + \frac{2}{r} \partial_r - \kappa^2 \mright) p_<(r) = 0
\end{equation}
with the inverse length $\kappa := \sqrt{\lambda/D}$, describing the penetration depth into the reactive
domain.
The flux takes the form $j_<(r) = -D \partial_r p_<(r)$, which turns the boundary conditions for the flux, \cref{eq:lower-bc-flux,eq:cont-bc-flux}, into von Neumann conditions for the concentration, $p_<'(0) = 0$ and $p_<'(R) = K/4\pi D R^2$.
\Cref{eq:reaction-diffusion-inner} is equivalent to
$\mleft(\partial_r^2  - \kappa^2 \mright) [r p_<(r)] = 0$, and the boundary value problem is solved by \cite{Erban2009}
\begin{equation}
  \label{eq:densityErbChap}
  p_<(r) = \gamma \,\frac{\sinh (\kappa r)}{\kappa r}
\end{equation}
with the constant $\gamma$ fixed by the upper boundary;
in particular, $\gamma$ is proportional to the reaction frequency~$K$.
Matching inner and outer solutions for $p(r)$, \cref{eq:outerSolutionFree,eq:densityErbChap},
at the interface, $r=R$, leads to $\gamma = c_A / \cosh(\kappa R)$, and Doi's result for the reaction rate constant \cite{Doi1975,Erban2009} follows:
\begin{align}
  k = 4\pi D R \mleft [ 1 - \frac{\tanh (\kappa R) }{\kappa R} \mright ].
  \label{eq:rateErbChap}
\end{align}

The solution naturally decomposes as in \cref{eq:total-rate-constant} into Smoluchowski's encounter rate
$k_\mathrm{e} = 4\pi D R$, see \cref{eq:DebyeRate}, and a formation rate
\begin{equation}
    \label{eq:potential-free-formation-rate}
    k_\mathrm{f} = 4\pi D R\mleft[\kappa R \cotanh(\kappa R) - 1\mright],
\end{equation}
with $\cotanh(x)=1/\tanh(x)$.
In the fast-diffusion limit, $\kappa R\ll 1$, i.e., when the reaction propensity $\lambda$ is low, the formation rate
$k_\mathrm{f} \simeq (4\pi/3) R^3 \lambda$ is simply the product of the reaction volume $V_R = (4\pi/3) R^3$ and the propensity,
reflecting well-mixed conditions inside the reaction volume ($p_<(r) = \mathit{const}$).
For fast reactions, $\kappa R\gg 1$, we obtain
$k_\mathrm{f} \simeq 4\pi R^2\kappa^{-1}\lambda$,
which we interpret as reactions being restricted to a volume $4\pi R^2 \kappa^{-1}$, that is a thin shell of radius $R$ and width $\kappa^{-1}$.

\section{Reaction rates and spatial distributions in the presence of an interaction potential}
\label{sec:general-solution}

For the general solution to the reaction--diffusion problem, \cref{eq:radialSmoluchowski,eq:radialCurrent}, in the presence of an interaction potential, it remains to find a solution inside the reaction radius (inner domain) and to match it with \cref{eq:outerSolution}.
As boundary condition we use $j_<(0) = 0$, \cref{eq:lower-bc-flux}, and solve for the current $j_<(r)$ first.

\subsection{Constant potential inside the reaction volume}

As a preliminary to the general discussion, we consider the analytically accessible situation that the interaction potential
is constant within the reaction volume, i.e., $U(r) = U(R)$ for $r \leq R$.
This may be useful in modelling reactions in electrolytes while neglecting excluded volume effects.
Then the inner solution equals the non-interacting case, \cref{eq:densityErbChap}, and can be matched with \cref{eq:outerSolution} to find the reaction rate constant
\begin{equation}
    k = 4\pi D \left( \frac{R \, g(R)}{\kappa R \cotanh (\kappa R) - 1} + \int _R ^\infty g(r) \,\d r \right)^{-1}.
    \label{eq:macroscopicRateConstantPotential}
\end{equation}
In particular, the encounter rate $k_\mathrm{e}$ is equal to Debye's result, \cref{eq:DebyeRate}, whereas the formation rate is suppressed by a factor $R^2 g(R) = \e^{\beta U(R)}$ relative to the non-interacting value, \cref{eq:potential-free-formation-rate},
and the total rate is the harmonic mean of both, \cref{eq:total-rate-constant}.

\subsection{Solution for arbitrary potentials}
\label{sec:innerSolutionBounded}

We proceed along the lines of the potential-free case, \cref{sec:innerSolutionWithout}, and solve
\cref{eq:radialSmoluchowski,eq:radialCurrent} inside the reaction volume, $0 \leq r \leq R$, subject to the boundary conditions
\cref{eq:lower-bc-flux,eq:cont-bc-flux}.
Applying the differential operator
$\e^{-\beta U(r)}\partial_{r} \e^{\beta U(r)}$ on both sides of \cref{eq:radialSmoluchowski}
and identifying the flux on the right hand side, one finds the following Dirichlet problem for the dimensionless function
$\psi(r):=-4\pi r^2 j_<(r) / K$:
\begin{subequations}
\label{eq:bvp-psi}
\begin{gather}
  \psi''(r) + \left( \beta U'(r) - \frac{2}{r}  \right) \psi'(r) - \kappa^2 \psi(r) = 0 \,,
    \label{eq:bvp-psi-a} \\
  \psi(0) = 0\,, \quad \text{and} \quad \psi(R) = 1 \,.
\end{gather}
\end{subequations}
In the absence of an explicit solution, we use the method of finite differences\cite{Smith1985numerical} to compute, in particular, the
derivative on the reaction boundary, $\psi'(R)$.
The latter determines the concentration on the boundary via \cref{eq:radialSmoluchowski}:
\begin{equation}
  p_<(R) = \psi'(R) K / 4 \pi R^2 \lambda \,.
  \label{eq:conc-boundary}
\end{equation}

Eventually, the reaction frequency $K$ is obtained by matching inner and outer solutions for the concentration, \cref{eq:cont-bc-conc}.
Employing the numerical value for $\psi'(R)$ and our previous result, \cref{eq:outerSolution}, we have
\begin{equation}
  \frac{K}{4 \pi R^2 \lambda} \,\psi'(R) = \e^{-\beta U(R)} \mleft[ c_{A} - \frac{K}{4\pi D} \int _{R} ^{\infty} g(s) \, \d s \mright ].
\end{equation}
Solving for $K=k/c_A$, yields an exact, closed expression for the macroscopic rate constant $k$, which is one of our main results:
\begin{equation}
    \label{eq:macroscopicRate}
    k = 4\pi D \left[ \int_{R}^{\infty} g(s) \d s + \frac{g(R) \,\psi'(R)}{\kappa^{2}} \right]^{-1} \,;
\end{equation}
the pair potential enters through the function $g(r) := \e^{\beta U(r)} r^{-2}$.
The result naturally displays the decomposition of \cref{eq:total-rate-constant}, and we identify the formation rate as
\begin{equation}
    \label{eq:formationRate}
    k_\mathrm{f} = \frac{4\pi \lambda}{g(R) \, \psi'(R)} \,,
\end{equation}
which appears to be proportional to the reaction propensity $\lambda$;
in fact, the value of $\psi'(R)$, as given by Eqs.~\eqref{eq:bvp-psi}, indirectly depends on $\lambda$ as well.
Noteworthy, the diffusion-limited encounter rate $k_\mathrm{e}$ is the same as for the Debye problem, see \cref{eq:DebyeRate},
and the classical result, $k = k_\mathrm{e}$, is recovered in the limit of instantaneous reactions, $\lambda \to \infty$, i.e., for vanishing $k_\mathrm{f}^{-1}$.

An alternative expression for the formation rate $k_\mathrm{f}$ in terms of the concentration $p(R)$ is obtained by substituting $\psi'(R)$ using \cref{eq:conc-boundary} and $K=k c_A$, which yields
$k_\mathrm{f} = k c_A \e^{-\beta U(R)} / p(R)$.
Employing the decomposition of the total rate $k$ [\cref{eq:total-rate-constant}] and solving for $k_\mathrm{f}$, one finds
\begin{equation}
  k_\mathrm{f} = k_\mathrm{e} \left[ \frac{c_A \e^{-\beta U(R)}}{p(R)} - 1 \right].
  \label{eq:formationRate2}
\end{equation}
Interestingly, the formation rate is fully specified by the encounter rate $k_\mathrm{e}$ and the concentration at the reaction boundary relative to its equilibrium value.
However, the computation of $p(R)$ requires the full solution of the reaction--diffusion problem.

The concentration profile $p(r)$ follows from integration of \cref{eq:radialCurrent} in terms of $\psi (r)$ and using continuity, \cref{eq:cont-bc-conc}, to eliminate $p_<(R)$ to find
\begin{align}
    \label{eq:density-inner}
    p(r) = c_A \e^{-\beta U(r)} \left[1  - \frac{k}{4 \pi D} \int_r ^{\infty} g(s) \psi(s) \d s \right ],
\end{align}
with the convention $\psi(r) = 1$ for $r>R$. Alternatively the density profile can also be found by
\cref{eq:radialSmoluchowski}, from the solution $\psi(r)$ as $p_<(r) = \psi'(r) K / 4 \pi r^2 \lambda$. However, we observed the numerical integration in \cref{eq:density-inner} to yield smaller
errors.

\subsection{Perturbative solution for slow reactions}
\label{sec:perturbation}

Slow reactions, $\lambda \ll D R^2$, corresponding to a well-mixed reaction volume, are described by a large penetration depth $\kappa^{-1} \gg R$. This suggests to expand the concentration profile $p_<(r)$ in the small parameter $\kappa R \ll 1$, introducing functions $p_0, p_1, \dots$:
\begin{equation}
    \label{eq:perturbation-ansatz}
    p_<(r) = p_0(r) + (\kappa R)^2 p_1(r) + O\bigl((\kappa R)^4\bigr) \,;
\end{equation}
here, we neglect terms of order $(\kappa R)^4$.
Corresponding fluxes $j_0(r), j_1(r), \dots$ are defined by virtue of \cref{eq:radialCurrent}.
Inserting the expansion into \cref{eq:radialSmoluchowski} for $r\leq R$ and sorting by powers of $\kappa^2=\lambda/D$, one finds that the 0\textsuperscript{th} order is satisfied by the equilibrium distribution in the absence of reactions:
\begin{equation}
    \label{eq:perturbation-zero-order}
    p_0(r) = c_A \e ^{-\beta U(r)} \,,
\end{equation}
which is accompanied by a vanishing flux, $j_0(r) \equiv 0$, due to detailed balance.
The flux $j_1(r)$ at order $(\kappa R)^2$ obeys
\begin{equation}
  \frac{1}{r^2} \, \partial_r r^2 j_1(r) = -\kappa^2 D \, p_0(r) \,,
\end{equation}
which can be integrated to yield
\begin{equation}
  j_1(r) = -\frac{\kappa^2 D c_A}{r^2} \int_{0}^{r}\e^{-\beta U(s)}s^2 ds
\end{equation}
for $0 \leq r \leq R$,
where we used the boundary condition $j(0) = 0$ [\cref{eq:lower-bc-flux}].
With this, the reaction rate constant $k$ follows from \cref{eq:relation-k-and-flux} straightforwardly:
\begin{equation}
  \label{eq:macroscopicRatePerturbative}
  k = \kappa^2 D \int_{0}^{R} \e^{-\beta U(r)} \, 4\pi r^2 \d r + O\bigl((\kappa R)^4\bigr) \,.
\end{equation}
It allows for a simple interpretation valid for slow reactions:
the macroscopic rate $k \simeq \lambda V_\mathrm{eff}$
is the product of the reaction propensity $\lambda$ and an effectively accessible reaction volume \cite{frohner2018reversible},
\begin{equation}
  \label{eq:Veff}
  V_\mathrm{eff} = \int_{|\vec r| \leq R} \e^{-\beta U(\vec r)} \, \d^3 r \,.
\end{equation}

\subsection{Numerical details}
\label{sec:numerics}

\begin{figure}
    \includegraphics[width=\figwidth]{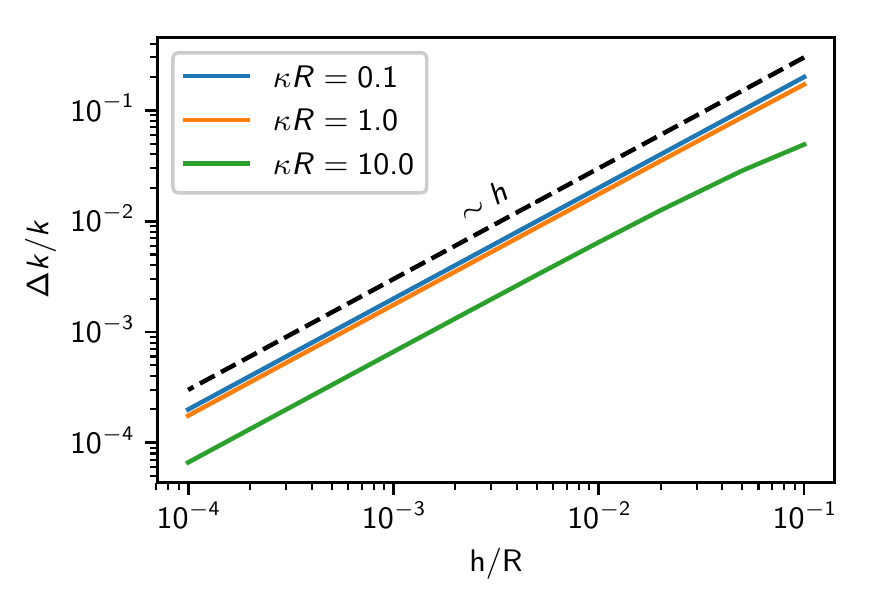}
    \caption{Relative error $\Delta k/k$ of the reaction rate constant $k$ of the numerical solution [\cref{eq:macroscopicRate}] with respect to the analytical solution [\cref{eq:rate-logpotential}] for a diverging potential [\cref{eq:logpotential}]. The numerical result is obtained for different discretisation widths $h$ given in units of the reaction radius $R$ and for different reactivities $\kappa R$. The dashed line depicts a linear scaling,
    $\Delta k/k \sim h$.
    }
    \label{fig:numerics_test}
\end{figure}

The computation of the reaction rate [\cref{eq:macroscopicRate}] for arbitrary potentials and reaction parameters requires the numerical solution of the boundary-value problem, \cref{eq:bvp-psi}, and of the integral, \cref{eq:DebyeRate}.
We checked our numerical implementation by comparing to the analytically exactly tractable, albeit peculiar case of a logarithmic potential,
\begin{equation}
    \label{eq:logpotential}
    U(r) = \begin{cases} - 2 k_B T \log(r/R), & r < R \\ 0,  & \text{otherwise.} \end{cases}
\end{equation}
With this, $g(r) = R^{-2} \,\theta(R - r)$ is a step function, and the coefficient $\beta U'(r) - 2/r$ in \cref{eq:bvp-psi-a} reduces to $-4/r$.
The differential equation can be solved using computer algebra, yielding $\psi'(R)$ and the reaction rate according to \cref{eq:macroscopicRate} as
\begin{equation}
    \label{eq:rate-logpotential}
    k = 2 \pi D R \left\{ 3 - \frac{(\kappa R)^2}{(\kappa R)^2 - 2[\kappa R \coth (\kappa R)-1]}  \right\}.
\end{equation}

The Debye rate was computed via the adaptive quadrature routines from QUADPACK.
For numerical solutions to \cref{eq:bvp-psi}, we used the method of finite differences \cite{Smith1985numerical} by discretising the domain $[0, R]$ into $N$ sub-intervals of equal size $h:=R/N$.
Let us note that at the outer most grid points, $r=0$ and $r=R$, \cref{eq:bvp-psi-a} does not require evaluation if central differences are used to compute $\psi'(r)$ and $\psi''(r)$ from $\psi(r)$.
For a range of values of $\kappa R$, we computed the error $\Delta k$ between the numerical and the analytical results for the rate, see \cref{fig:numerics_test}.
The relative error $\Delta k/k$ scales approximately linearly with $h$ and decreases with increasing $\kappa R$.
For the worst case studied, $\kappa R = 0.1$, we conclude that an accuracy better than $10^{-3}$ is reached by choosing a grid spacing of $h = 10^{-4} R$, which is still well feasible in terms of computational costs.
This value of $h$ is used for all subsequent calculations.

Finally, we have checked that all terms in \cref{eq:bvp-psi-a} are bounded.
In particular, we argue that the term $[\beta U'(r) - 2/r] \psi'(r)$ vanishes in the limit $r\to 0$.
The expression is proportional to $[\beta U'(r) - 2/r] r^2 p(r)$ after re-substituting $\psi(r)$
and using \cref{eq:radialSmoluchowski}.
Further, we anticipate that the concentration profile is bounded from above by the equilibrium distribution,
$p(r) \leq c_A \e^{-\beta U(r)}$, as reactions can only lower the concentration in the reaction volume, see \cref{fig:profiles}.
With this, $(2/r) \, r^2 p(r) \to 0$ and
$|\beta U'(r)\,p(r)| \leq c_A \left| \partial_r \e^{-\beta U(r)} \right|$,
and it remains to show that
$\left| \partial_r \e^{-\beta U(r)} \right| \xrightarrow{r \to 0} 0.$
This is fulfilled by certain logarithmic potentials, such as in \cref{eq:logpotential}, and by algebraically diverging potentials, $\beta U(r \to 0) \simeq a r^{-m}$ with $a,m > 0$.
In the latter case, putting $y := r^{-m}$ we have
$\left| \partial_r \e^{-\beta U(r)} \right| \simeq a m \, y^{(m+1)/m} \e^{-a y} \to 0$ as $y \to \infty$.

\section{\protect\iPRD{} simulations}
\label{sec:simulations}

Complementary to the preceding theoretical analysis, we have performed extensive simulations of the microscopic reaction--diffusion dynamics in the steady state.
We ``measure'' the absolute reaction rate $k$ of the reaction \eqref{eq:the-reaction} and the radial distribution function $p(r)$ of \ce{A} molecules relative to a \ce{B} molecule.

\subsection{Simulation setup and protocol}

Stochastic simulations of the interacting particle-based reaction--diffusion dynamics (iPRD) are performed with the software ReaDDy~2 \cite{Hoffmann2018,Schoneberg2013}, which integrates the motion of particles and reactions between them explicitly in three-dimensional space.
In ReaDDy, time is discretised into steps of fixed size $\Delta t$. A single step consists of first integrating the Brownian motion of molecules via the Euler--Maruyama scheme and then handling reaction events according to the Doi model (\cref{sec:model}).
After each step, one can evaluate observables, such as the positions of particles or the number of reactions that occurred.

The simulation setup is constructed spherically symmetric around a single \ce{B} molecule in the coordinate origin, as depicted in \cref{fig:scheme}.
In particular, we use a spherical domain of finite radius $L$, which will be filled with \ce{A} molecules such that at the boundary, $r=L$, the concentration $p(L)$ of \ce{A} molecules matches a given constant.
Within the whole domain, \ce{A} particles diffuse subject to the interaction potential $U(r)$, whereas the \ce{B} molecule is fixed in space; here, we restrict ourselves to potentials that are cut off at a distance $r_c < L$.
The conversion reaction \eqref{eq:the-reaction} takes place with reaction propensity $\lambda$ inside the sphere with $r \leq R$.
We have run a large number of simulations for varying propensity $\lambda$ and different potentials $U(r)$, see below.
Simulation units were chosen such that distances are measured in terms of the reaction radius $R$, energies in terms of the thermal energy $k_BT$, and times in terms of the combination $\tau_\mathrm{d} := R^2/D$, which is proportional to the time to explore the reaction volume by diffusion.
The parameters used are listed in \cref{tab:parameters}, in particular, a time step $\Delta t=10^{-4} \tau_\mathrm{d}$ was used throughout production runs.
\footnote{The chosen time step is sufficiently small to be suitable for the Lennard-Jones potential, which generally calls for much smaller integration steps than the harmonic repulsion due to an increased stiffness.}

\begin{table}
    \centering %
    \begin{tabular}{lcll}
        \toprule
        Quantity  & Symbol  & Value & Unit \tabularnewline
        \midrule
        Propensity of reaction \eqref{eq:the-reaction} & $\lambda$ & varies & $\tau_\mathrm{d}^{-1}$ \tabularnewline
        Soft repulsion strength & $b$ & $40$ & $k_BT/R^2$ \tabularnewline
        Soft repulsion range & $r_0$ & $1$ & $R$ \tabularnewline
        LJ interaction strength & $\epsilon$ & $1$&$ k_BT$ \tabularnewline
        LJ interaction range & $\sigma$ & $(26/7)^{-1/6}$&$R$ \tabularnewline
        LJ cutoff radius & $r_c$ & $2.5$&$R$ \tabularnewline
        \midrule
        Integration time step & $\Delta t$ & $10^{-4}$&$\tau_\mathrm{d}$\tabularnewline
        Radius of simulation domain & $L$ & $10$&$R$\tabularnewline
        Width of factory shell & $\Delta L$ & $5$&$R$\tabularnewline
        Number of factory particles & $N_f$ & $1.5\times 10^4$ & $1$ \tabularnewline
        Propensity to create \ce{A} & $f_+$ & $0.01$&$\tau_\mathrm{d}^{-1}$\tabularnewline
        Propensity to absorb \ce{A} & $f_-$ & $0.01$&$\tau_\mathrm{d}^{-1}$\tabularnewline
        \bottomrule
    \end{tabular}
    \caption{Parameters used in the particle simulations.
        Basic units of length, time, and energy are $R$, $\tau_\mathrm{d} := R^2/D$, and $k_BT$, respectively.
    }
    \label{tab:parameters}
\end{table}

Aiming at the simulation of a stationary reaction kinetics, we coat the domain by a \emph{factory shell},
with radial coordinates in $r\in[L,L+\Delta L]$, that yields a constant supply of \ce{A} molecules.
Adjacent to the shell, for $r \geq L + \Delta L$, an external harmonic potential is added that prevents \ce{A} molecules from escaping and thereby closing the simulation domain.
The factory shell contains $N_f$ factory (\ce{F}) particles, which are fixed in space at random positions according to a uniform distribution.
\ce{F} particles create and absorb \ce{A} molecules through the reversible reaction
\begin{equation}
    \label{eq:factory-reactions}
    \ce{F <=>[$f_+$][$f_-$] F + A} \,.
\end{equation}
The forward reaction has propensity $f_+$ and is of fission type: a new \ce{A} molecule is placed at a random distance $d\in [0,R_f]$ from the active \ce{F} particle.
The backward reaction is of fusion type, by which an \ce{A} molecule is absorbed with propensity $f_-$ if it is closer than $R_f$ to an \ce{F} particle.
Due to the fact that the number of \ce{F} particles is conserved, the factory reactions \eqref{eq:factory-reactions} are pseudo-unimolecular, i.e. they can be reduced to
\begin{equation}
    \label{eq:factory-pseudo-reactions}
    \ce{A <=> $\varnothing$},
\end{equation}
which leads to a steady-state concentration $p(L)$ of \ce{A}s.
The latter depends also on the outflux $K = 4 \pi L^2 |j(L)|$ of \ce{A} molecules,
which can diffuse freely into and out of this shell and migrate towards the origin due to the reaction of interest, \cref{eq:the-reaction}.
Lacking an \emph{a priori} knowledge of the concentration $p(L)$ and the concentration $c_A$ in the far field ($r\to \infty$), we run simulations with a certain set of parameters
$N_f$, $f_+$, $f_-$, and $R_f$ and estimate the resulting value of $c_A$ accurately from the observed steady-state profile $p(r)$.
Specifically, we fit the solution
$p(r) = c_A - K / 4\pi D r$ [\cref{eq:outerSolutionFree}],
to the data for $p(r)$ in the range $\max(R, r_c) \leq r \leq L$, where both interactions and reactions are absent and \ce{A} molecules diffuse freely.
This yields the extrapolated concentration at far distances, $p(r \to \infty) = c_A$.
Note that the reaction frequency $K$ is directly available from the simulation by counting reaction events.

The above procedure relies on the fact that shifting the upper boundary from infinity to $r = L$ merely shifts the concentration $p(r)$ by an additive constant, leaving the integral flux through spheres of radius $r$ unchanged, provided that $r$ is outside of the interaction range.
This is a consequence of Gauss's theorem, see also \cref{eq:radialCurrent}.
Therefore, simulation results with a finite volume can be mapped exactly to the infinite case upon using
the effective far-field concentration $c_A$ as determined above.

A data production cycle starts with uniformly distributing \ce{A} molecules in the factory shell with a concentration that roughly anticipates the expected $c_A$.
This initial state is relaxed by evolving the reaction--diffusion dynamics for a time span of $t_\mathrm{eq}=300 \tau_\mathrm{d}$, by executing $3\times 10^5$ integration steps with a coarser time step size of $\Delta t=10^{-3} \tau_\mathrm{d}$. Equilibration is verified by observing that the number of \ce{A} particles does not vary significantly.
The time step is then decreased to $\Delta t=10^{-4} \tau_\mathrm{d}$ and the system equilibrated for another time span of $30 \tau_\mathrm{d}$.
During the subsequent production run of length similar to $t_\mathrm{eq}$, we record the two main observables:
\begin{enumerate*}
  \item the concentration profile $p(r)$ as the radial distribution function (RDF) of \ce{A} molecules relative to the \ce{B} molecule in the centre, and
  \item the number of reactions \eqref{eq:the-reaction} that were performed in each integration step, yielding the reaction frequency $K$ and thus the macroscopic reaction rate constant $k=K/c_A$.
\end{enumerate*}
Observing the RDF in the case without a reaction and comparing it against the Boltzmann distribution is used to verify the time step.

One such simulation procedure took roughly 512 hours on a single CPU. Simulations were run for 3 different potentials and 5 different propensities, for each combination statistical averages over 13 independent realisations were taken, altogether yielding 195 simulations that were run in parallel.
The cumulative CPU time amounts to 100{,}000 hours.

\subsection{Pair potentials}
\label{sec:potentials}

\begin{figure}
    \includegraphics[width=\figwidth]{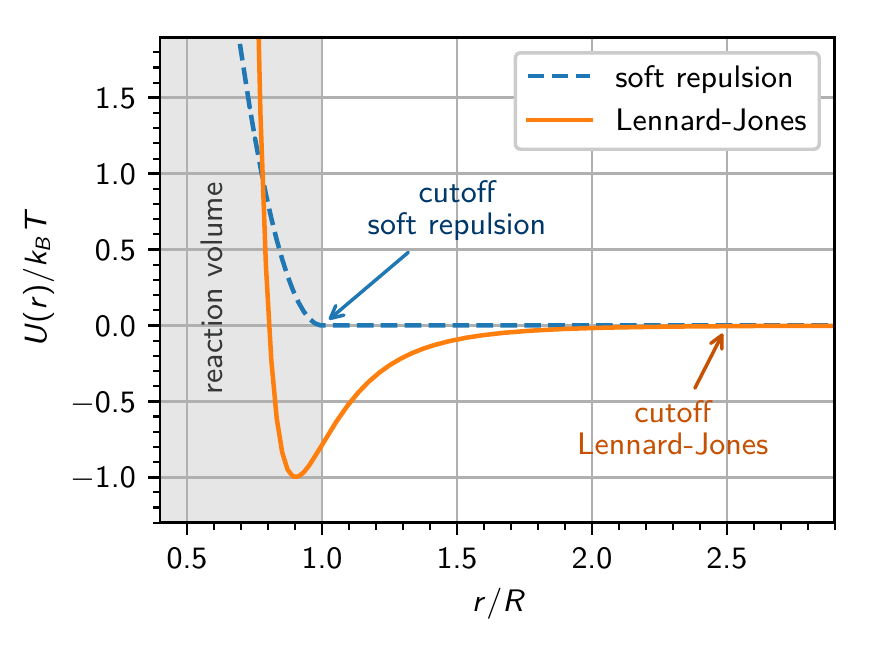}
    \caption[Pair potentials]{Pair potentials $U(r)$ used in our study of the steady-state reaction kinetics
    [\cref{eq:harmonic-repulsion,eq:lennard-jones}]
    for the parameters given in \cref{tab:parameters}.
    The separation $r$ of molecule centres is given in units of the reaction radius $R$,
    and the potential energy $U$ is given in terms of the thermal energy $k_BT$;
    the shaded region marks the reaction sphere in which reaction \eqref{eq:the-reaction} can occur.
    Arrows indicate the location of the interaction cutoffs.}
    \label{fig:potentials}
\end{figure}

In the following, we consider two different isotropic pair potentials for the interaction between \ce{A} and \ce{B} molecules, and we compare to the non-interacting case ($U=0$).
The employed potentials are visualized in \cref{fig:potentials}, and all relevant parameters are given in \cref{tab:parameters}.
The first potential describes an ultra-soft steric repulsion, which is common for macromolecules such as polymer rings \cite{Poier2015anisotropic}.
For simplicity, we assume that \ce{A} and \ce{B} molecules repel each other only when their centres are within a cutoff radius $r_0$,
and we use a harmonic form:
\begin{equation}
    \label{eq:harmonic-repulsion}
    U(r) = \frac{1}{2}b (r-r_{0})^{2} \,, \quad r\leq r_0 \,,
\end{equation}
and $U(r) = 0$ otherwise;
here, $b>0$ is a harmonic spring constant chosen to be stiff, $b r_0 \gg k_B T$, and we set the cutoff equal to the reaction radius, $r_0=R$.

The second potential is a commonly truncated form of the Lennard-Jones (LJ) potential, which combines a strong steric repulsion of nearly overlapping molecules with a short-range attraction due to van der Waals forces:
\begin{equation}
    U(r) = 4 \epsilon \left[ (\sigma/r)^{12} - (\sigma/r)^{6} \right] \theta(r_c - r),
    \label{eq:lennard-jones}
\end{equation}
with $\sigma$ and $\epsilon>0$ being a length and an energy, respectively, that set the range and the strength of the interaction.
The value of $\epsilon$ is also the depth of the potential well at $r=\sigma$.
Here we choose $\sigma$ such that the potential minimum lies \textit{within} the reaction volume, specifically, the inflection point of $U(r)$ is set at the boundary, $R = (26/7)^{1/6} \sigma \approx 1.24\sigma$.
The attractive part of the interaction is truncated at $r_c = 2.5 R$.

\section{Results and discussion}
\label{sec:discussion}

\subsection{Macroscopic rates}

\begin{figure*}
    \includegraphics[width=\linewidth]{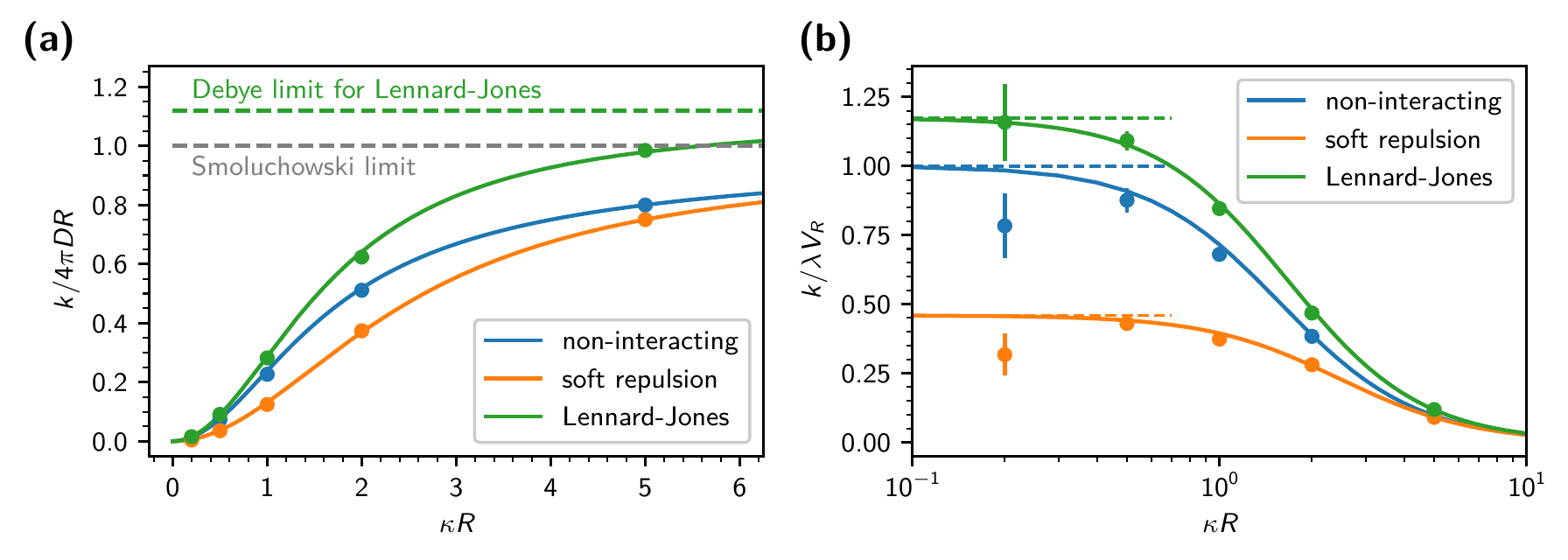}
    \caption{\textbf{(a):} Macroscopic rate constant $k$ as a function of the reactivity $\kappa R$ with the inverse penetration depth $\kappa=\sqrt{\lambda/D}$ and the reaction radius $R$ for different pair potentials $U(r)$.
    Data are given relative to the Smoluchowski rate constant $4\pi D R$ (grey dashed line) in terms of the relative diffusion constant $D = D_{\ce{A}} + D_{\ce{B}}$ and the reaction radius $R$.
    Symbols are results of interacting particle-based stochastic simulations of the reaction--diffusion process (iPRD simulations).
    Solid lines show theoretical predictions obtained from exact expressions [non-interacting case, \cref{eq:rateErbChap}]
    or quasi-analytic solutions [soft harmonic repulsion and LJ potential, \cref{eq:macroscopicRate}] of the reaction--diffusion problem, Eqs.~\eqref{eq:reaction-diffusion}.
    The green dashed line indicates the Debye limit, \cref{eq:DebyeRate}, for the LJ potential.
    \textbf{(b):} Macroscopic rate constant $k$ as a function of the reactivity $\kappa R$ normalized by the perturbative solution $k^{(0)} \simeq \lambda V_R$ of the non-interacting case for slow reactions
    [\cref{eq:macroscopicRatePerturbative}].
    Dashed lines indicate the ratios of the accessible to the total reaction volume $V_\mathrm{eff}/V_R$ for each potential [\cref{eq:Veff}], which is the prediction of perturbation theory.
    }
    \label{fig:rates}
\end{figure*}

Simulation results for the reaction rate constant $k$ as a function of the propensity $\lambda=\kappa^2 D$ are shown in \cref{fig:rates} for the above potentials.
They are compared to the theoretical predictions from the reaction--diffusion problem, Eqs.~\eqref{eq:reaction-diffusion}, as follows:
For the non-interacting case ($U=0$), the exact solution is available in closed form, \cref{eq:rateErbChap}.
For the soft repulsion and the LJ potential, the solution is available only in quasi-analytic form, \cref{eq:macroscopicRate}, i.e., the final expressions for $k$ are explicit in terms of a numerical quadrature as in the Debye problem and the numerical solution to a one-dimensional boundary value problem in the interior of the reaction sphere, see \cref{sec:numerics}.
As dimensionless control parameter we choose the combination $\kappa R = R \sqrt{\lambda/D}$, which distinguishes the reaction- and diffusion-limited regimes, $\kappa R \ll 1$ and $\kappa R \gg 1$, respectively.
Equivalently, $(\kappa R)^2 = \lambda \tau_\mathrm{d}$ controls the reaction propensity relative to the diffusion time $\tau_\mathrm{d} = R^2/D$.

For all choices of the potential, the agreement between theory and simulations is excellent, see \cref{fig:rates}a.
In all three cases, the reaction rate $k$ increases monotonically with the reaction propensity $\lambda$ and saturates at Debye's result, \cref{eq:DebyeRate}, for a diffusion-limited reaction ($\kappa R \to \infty$).
In this limit, the reaction occurs almost surely upon first contact and details inside of the reaction volume become irrelevant, the formation rate diverges, $k_\mathrm{f} \to \infty$.
Note that for the truncated soft repulsion, \cref{eq:harmonic-repulsion}, the limiting value equals the Smoluchowski rate as the potential is zero in the outer domain.
For slow reactions, $\kappa R \ll 1$, the initial increase of $k$ depends quadratically on $\kappa R$ and it coincides with the prediction $k \simeq \lambda V_\mathrm{eff}$ of perturbation theory, \cref{eq:macroscopicRatePerturbative}.
This regime is better visualised by normalising $k$ with the perturbation result for the non-interacting case, $k^{(0)} = \lambda V_R$, where $V_R = (4\pi/3)R^3$, see \cref{fig:rates}b.
From the limit $\kappa R \to 0$ it is evident that also the constant of proportionality $V_\mathrm{eff}$ as calculated from \cref{eq:Veff} matches very well with the numerical results.
For $\kappa R=0.2$ noticeable relative deviations are seen in the simulation data,
indicating that the slow-reaction regime is challenging to explore by the particle-based approaches such as iPRD.
The figure shows further that the perturbation solution deviates by no less than 10\% from the full solution for $\kappa R \lesssim 0.5$.

How is the reaction rate constant $k$ changed due to the presence of the investigated potentials?
A repulsion within the reaction volume slows down the reaction relative to the non-interacting case, which we attribute to the greatly diminished accessible reaction volume (\cref{fig:rates}, soft repulsion).
The effect is most pronounced for slow reactions, which are most sensitive to a reduction of the actual penetration depth
relative to its value $\kappa^{-1}$ of the free case.
Evaluating \cref{eq:Veff} for the specific harmonic repulsion used here, $V_\mathrm{eff}$ and thus $k$ are reduced by a factor of $\approx 2.2$ relative to the non-interacting case.

\begin{figure}
  \includegraphics[width=\figwidth]{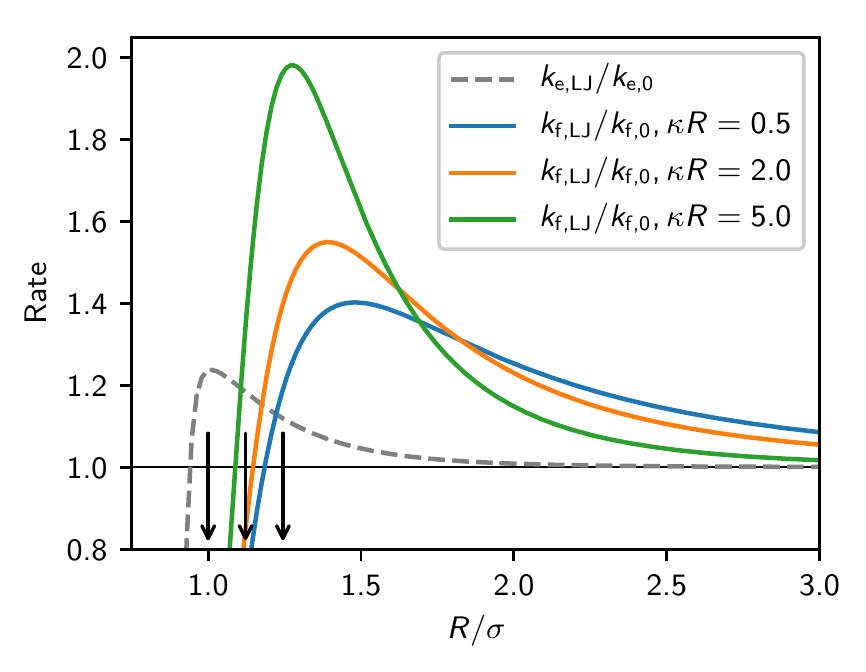}
  \caption{Dependence of the partial reaction rates $k_e$ and $k_f$ on the attractive part of the LJ potential with depth $\epsilon/k_BT=1$, which is tested by varying the interaction range $\sigma$ for fixed reaction radius $R$. The rates are normalised by their values for the non-interacting case, \cref{eq:DebyeRate,eq:potential-free-formation-rate}. Black arrows indicate the zero crossing, the minimum, and the inflection point of the Lennard-Jones potential.}
  \label{fig:ratesEnhancement}
\end{figure}

An attractive interaction between \ce{A} and \ce{B} molecules, on the other hand, is expected to enhance the encounter rate $k_\mathrm{e}$ and thus to speed up the overall reaction.
Already the short-ranged well of the truncated LJ potential, \cref{eq:lennard-jones}, suffices to increase $k_\mathrm{e}$ by 12\% with respect to the free case, \cref{eq:DebyeRate}.
Noting that only the part of the potential outside of the reaction volume, $r > R$, contributes to $k_\mathrm{e}$, we can test the dependence on the attraction by varying the interaction range $\sigma$ at fixed $R$, see \cref{fig:ratesEnhancement}.
The encounter rate becomes maximal at $\sigma = R$, i.e., when the integral in \cref{eq:DebyeRate}
is taken over the full domain where the potential is negative, $U(r) < U(r \to \infty)$.

The ramifications of the potential on the formation rate $k_\mathrm{f}$ are more subtle:
the strongly repulsive part of the LJ potential should lead to a decrease as the accessible reaction volume is diminished.
Concomitantly, the potential well induces an enrichment of \ce{A} molecules at the boundary of the reaction volume, which would increase $k_\mathrm{f}$.
The combination of both can lead to a non-monotonic dependence of the formation rate on the position of the reaction boundary relative to the potential well,
which indeed we observe in the numerical solutions to \cref{eq:formationRate}, see \cref{fig:ratesEnhancement}.
The position of the maximum in $k_\mathrm{f}$ depends on $\kappa R$ and shifts towards larger $\sigma/R$ for higher reaction propensity.
For the parameters given in \cref{tab:parameters}, the effectively accessible reaction volume is \emph{increased} by $\approx17\%$ over the free volume $V_R$ (\cref{fig:rates}b), and
for all $\kappa R$ the overall rate constant $k$ is larger than for non-interacting molecules.

\begin{figure}
  \includegraphics[width=\figwidth]{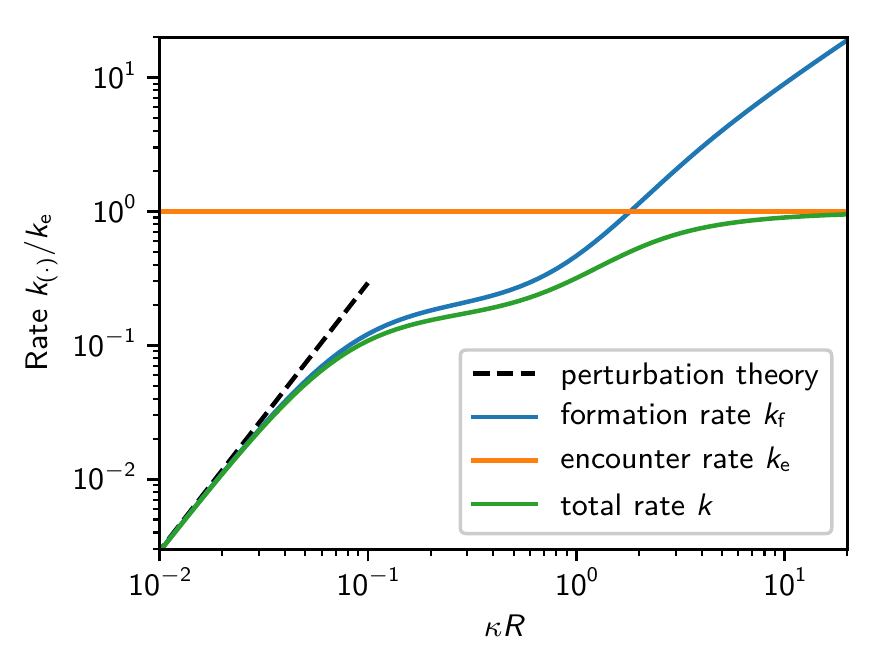}
  \caption{Encounter, formation and total rate constants as a function of the reactivity $\kappa R$ by changing the propensity $\lambda=\kappa^2 D$ for a Lennard-Jones potential with energy $\epsilon/k_BT=13$ and reaction radius $\sigma/R=0.1$. The dashed line shows the perturbative solution where $k\propto \kappa^2$.}
  \label{fig:partialRates}
\end{figure}

By the Markov property of the microscopic reaction--diffusion process, the total reaction rate constant $k$ is the harmonic mean of the partial rates for encounter and formation, \cref{eq:total-rate-constant}, and thus, $k$ is bounded from above by the smaller rate:
$k \leq \min(k_\mathrm{e}, k_\mathrm{f})$.
The relative importance of both processes depends on the rescaled reaction propensity $\kappa R$, which is nicely seen from \cref{fig:partialRates} for the Lennard-Jones potential with $\sigma/R=0.1$ and $\epsilon/k_BT=13$.
One reads off that the formation and diffusion-limited regimes, where the other contribution can safely be neglected, are delimited by $\kappa R \lesssim 10^{-1}$ and $\kappa R \gtrsim 10^1$, respectively.
Inbetween, there is a wide window of propensities, where both processes enter the overall rate constant.
Here, an enhanced availability of reactants due to the deep potential well compensates a slower reaction propensity so that the formation rate displays an approximately plateau-like behaviour for $0.1 \lesssim \kappa R \lesssim 0.5$.
For sufficiently fast reactions, the accumulation disappears and $k_f$ starts increasing again towards its large $\kappa R$ behaviour, $k_\mathrm{f} \sim \kappa R$, which resembles the potential-free case as reactions are confined to a thin shell near $r=R$.
Note that $k_f$ is a monotonic function of $\kappa R$, which follows from \cref{eq:formationRate2} and anticipating the monotonic decrease of $p(R)$ as $\kappa R$ increases, see \cref{fig:profiles}.

\begin{figure}
    \includegraphics[width=\figwidth]{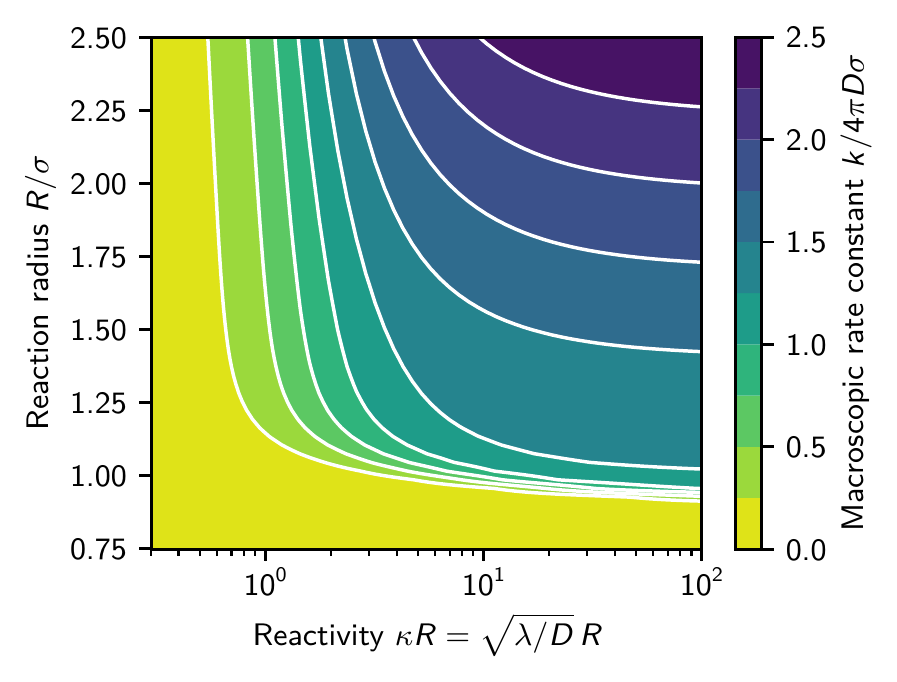}
    \caption{The macroscopic rate constant $k$ in the presence of a Lennard--Jones potential with particle diameter $\sigma$ and energy depth that is equal to the thermal energy $\epsilon=k_BT$. Here $k$ is a function of the unit--less reactivity $\kappa R=\sqrt{\lambda/D}\,R$ and a function of the reaction radius $R$, with the microscopic rate constant $\lambda$, relative diffusion constant $D$. $k$ is given in units of $4\pi D\sigma$, which is the encounter rate up to particle diameter if no reaction and potential would be present.}
    \label{fig:ratesMap}
\end{figure}

Motivated by the practical question how to choose the model parameters $\lambda$ and $R$ for given reaction rate $k$ and diffusivity $D$ and given interaction potential, we have scrutinized further the dependence of $k$ on both the propensity $\kappa R$ and the reaction radius $R/\sigma$, exemplified for the Lennard-Jones potential (\cref{fig:ratesMap}).
For slow reactions, $\kappa R \lesssim 1$, the rate constant $k$ is insensitive to the reaction radius.
In the diffusion-limited regime, $\kappa R \gtrsim 10$, the rate constant $k$ mainly depends on the reaction radius $R/\sigma$ and is insensitive to the value of $\kappa R$.
Inbetween, $1 \lesssim \kappa R \lesssim 10$, both parameters must be adjusted carefully.
From physical considerations, the reaction radius $R$ should be comparable to the molecular radius $\sigma$, which delimits the freedom in the choice of $\lambda$.

\subsection{Concentration profiles}

\begin{figure*}
    \centering
    \includegraphics[width=\linewidth]{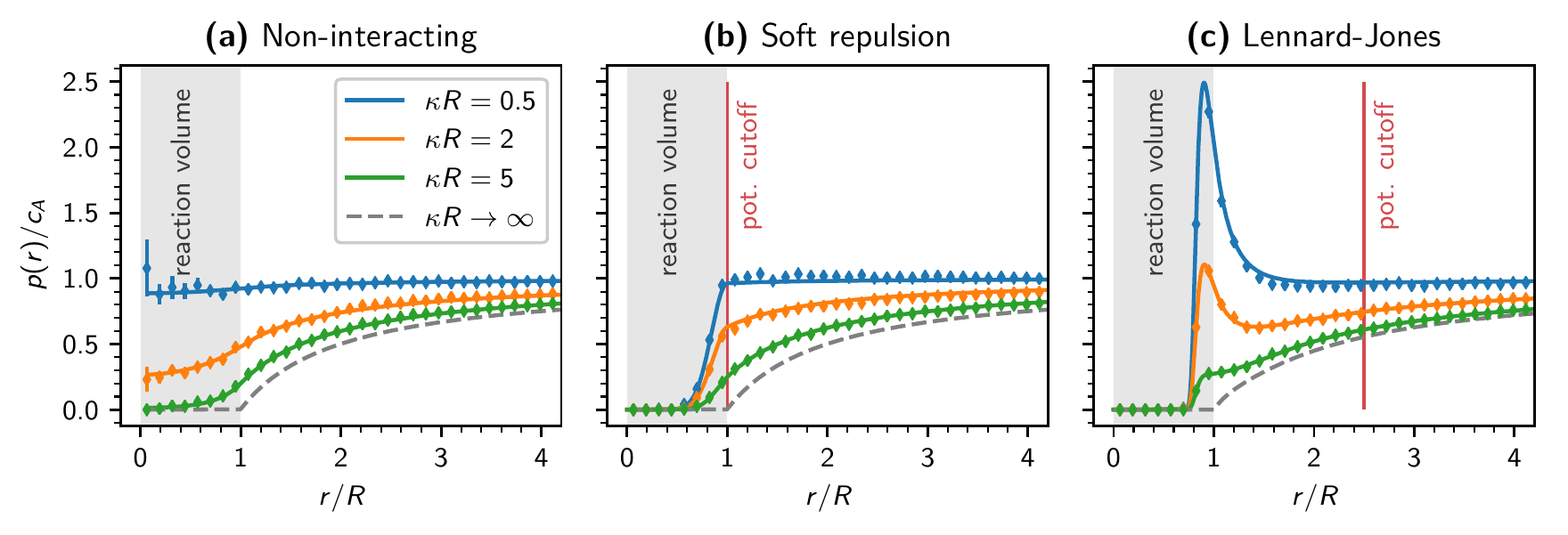}
    \caption{Radial distribution $p(r)$ of \ce{A} molecules around a \ce{B} molecule for different reaction propensities $\lambda$, here expressed by $\kappa=\sqrt{\lambda / D}$. The panels show results for (a) the non-interacting case,
    (b) the soft harmonic repulsion [\cref{eq:harmonic-repulsion}], and (c) a truncated LJ potential [\cref{eq:lennard-jones}].
    Data points are results from \iPRD\ simulations, and solid lines theoretical predictions from \cref{eq:densityErbChap,eq:outerSolutionFree} for the non-interacting case and from numerical solutions to \cref{eq:outerSolution,eq:bvp-psi} otherwise.
    Grey dashed lines represent the limit $\kappa R\to\infty$ of almost sure reactions upon contact [\cref{eq:outer-solution-debye}]. Grey shaded areas mark the interior of the reaction volume ($r \leq R$),
    and vertical lines indicate the respective positions $r_c$ of the potential cutoffs.
    }
    \label{fig:profiles}
\end{figure*}

Simulation results for the concentration profile $p(r)$, more precisely, the radial distribution of \ce{A} molecules relative to \ce{B}s,
are shown in \cref{fig:profiles} for three different propensities $\lambda$, expressed in terms of $\kappa=\sqrt{\lambda/D}$, and for the different interactions considered above.
The data are compared to the theoretical predictions developed in \cref{sec:solution-strategy,sec:general-solution},
and the quantitative agreement is very good for all cases studied.
Thus, the \iPRD\ simulations corroborate our theoretical analysis and the numerical results, which in turn are used to validate the implementation of the simulation algorithm.

For the non-interacting case (\cref{fig:profiles}a), we have closed analytic expressions for $p(r)$ inside and outside of the reaction volume, Eqs.~\eqref{eq:densityErbChap} and \eqref{eq:outerSolutionFree}, respectively.
For the soft repulsive and the LJ potentials [\cref{eq:harmonic-repulsion,eq:lennard-jones}], profiles in the outer domain are obtained from \cref{eq:outerSolution} by a quadrature, and in the inner domain from the numerical solution for $\psi'(r)$ of the boundary value problem, \cref{eq:bvp-psi}.
At distances $r > r_c$, where neither a reaction can occur nor a potential is present,
the constant flux implies for the profile, $p(r) = c_A (1 - k / 4\pi D r)$, see \cref{eq:outerSolutionFree}.

For slow reactions, $\kappa R \ll 1$, the concentration profile at leading order in $\kappa R$ is expected to equal the equilibrium distribution, $p_0(r) = c_A \e^{-\beta U(r)}$, subject to the specific boundary condition $p(r\to \infty) = c_A$ [\cref{eq:perturbation-zero-order}].
Indeed, for $\kappa R = 0.5$ both the numerical and simulation results for $p(r)$ are hardly distinguishable from $p_0(r)$ in all three cases studied, see \cref{fig:profiles}; for $U=0$ it holds $p_0(r) = c_A$ everywhere.
Upon increasing $\kappa R$, the concentration is decreasing uniformly and, in the limit of an instantaneous product formation, $\kappa R \to\infty$, the profile $p(r)$ vanishes inside the reaction volume and approaches Debye's solution, \cref{eq:outer-solution-debye}, outside as expected.
For the non-interacting case and the soft repulsive potential, the latter simplifies to Smoluchowski's result,
$p(r)=c_A(1 - R/r)$ for $r \geq R$; for the truncated LJ potential used here, the differences are small and hardly seen in the graph (\cref{fig:profiles}c).
Summarising, the equilibrium distribution and Debye's solution constitute upper and lower bounds on $p(r)$.

After having understood these limits, we will discuss the consequences of the interaction potential on the profiles in more detail.
Adding a soft repulsion within the reaction volume to mimic an excluded volume largely reduces the probability of finding a particle inside the reaction volume (\cref{fig:profiles}b) and thus suppresses the product formation rate $k_\mathrm{f}$ (see also \cref{fig:rates}b).
Yet, the effect is more pronounced for slow reactions as the interior of the reaction volume becomes less and less accessible upon increasing $\kappa R$, and we conclude that the repulsion is particularly relevant for slow reactions.
The attractive well of the LJ potential on the other hand induces an enrichment of \ce{A} molecules near the reaction boundary, which is more developed for smaller $\kappa R$ (\cref{fig:profiles}c).

\section{Conclusion}
\label{sec:conclusions}

We have studied the reaction kinetics of a bimolecular association process \ce{A + B -> X} in the steady state for molecules that diffuse in space and interact through an isotropic pair potential $U(r)$.
Within Doi's volume reaction model, we have calculated the reaction rate constant $k$ and the distribution function $p(r)$ of \ce{AB} pairs as a function of the microscopic reaction propensity $\lambda$.
The explicit dependence of the model on $\lambda$ enables us to systematically probe the kinetics from the well-mixed to the diffusion-limited regime.
The transition between the regimes is conveniently captured by the dimensionless quantity $R\sqrt{\lambda/D}$, which we abbreviate as the \emph{reactivity} $\kappa R$ of an \ce{AB} pair;
the length $\kappa^{-1}$ describes how far molecule centres can penetrate the reaction volume of radius $R$ before they react
and $D := D_{\ce{A}} + D_{\ce{B}}$ is the relative diffusion constant.
Specifically, our approach bridges between the two well-studied cases
$\kappa R \ll 1$ (reaction-limited or well-mixed) and
$\kappa R \gg 1$ (diffusion-limited or fast-reaction limit).
Similarly, $\lambda \tau_\mathrm{d}=(\kappa R)^2$ can be used to classify these regimes, however in terms of the residence time $\tau_\mathrm{d}=R^2/D$ in the reaction volume (as obtained for non-interacting molecules).

Over the entire spectrum of $\kappa R$ values and for arbitrary pair potentials, our analytical result for the reaction rate constant displays the Markovian decomposition
$k ^{-1} = k_\mathrm{e} ^{-1} + k_\mathrm{f}^{-1}$ into encounter $k_\mathrm{e}$ and formation $k_\mathrm{f}$ rates [\cref{eq:macroscopicRate}].
Thereby, $k_\mathrm{e}$ is always given by Debye's result \cref{eq:DebyeRate}.
Interestingly, $k_\mathrm{f}$ can be expressed in terms of $k_\mathrm{e}$ and the substrate concentration $p(R)$ at the reaction boundary, see \cref{eq:formationRate2}, the latter being non-trivial to calculate.
The well-mixed limit is dominated by the formation rate $k_\mathrm{f}$ and can be solved by perturbation theory (see \cref{sec:perturbation}), which yields $k = \lambda V_\mathrm{eff}$ in terms of the effectively accessible reaction volume $V_\mathrm{eff}$.
In the absence of a potential, $V_\mathrm{eff}$ simplifies to the volume of the reactive sphere $V_R=(4\pi/3) R^3$.
On the other hand, the diffusion limit is dominated by the encounter rate $k_\mathrm{e}$: a reaction occurs almost surely upon entering the reaction volume.
Our expression for $k$ reproduces the Smoluchowski encounter rate $4\pi D R$ in the absence of potentials and Debye's result \cite{Debye1942c}, when particles diffuse subject to an interaction potential $U(r)$.

In the application-relevant diffusion-influenced regime (see \cref{sec:general-solution}), where $k_\mathrm{e}$ is of comparable magnitude as $k_\mathrm{f}$,
we obtained semi-analytical expressions for the rate $k$ and the local concentration $p(r)$ that require numerical evaluation [\cref{eq:macroscopicRate,eq:density-inner}].
Practically, one has to solve a one-dimensional boundary value problem for the reaction--diffusion equation inside the reaction volume and to compute an integral over the domain outside the reaction volume;
the computational costs of both tasks are negligible.
We tested our numerical scheme against explicit analytic solutions for a logarithmically repulsive potential.
A closed expression for the rate $k$ is given for general potentials outside in the case that molecules do not interact if their centres are within the reaction volume [\cref{eq:macroscopicRateConstantPotential}]; this may be useful to model, e.g., reactions in electrolytes while neglecting excluded volume.

We have studied the detailed dependence of the rate $k$ on the reactivity parameter $\kappa R$ for two different potentials: a soft harmonic repulsion inside the reaction volume, and a truncated Lennard-Jones potential combining excluded volume and attraction.
Our numerical results for the rate $k$ and the concentration $p(r)$ show excellent agreement with extensive stochastic particle-based reaction-diffusion simulations.
We draw the following physical conclusions:
\begin{enumerate*}
\item A purely repulsive potential decreases both partial rates, $k_\mathrm{e}$ and $k_\mathrm{f}$, and so also the overall rate constant $k$ compared to the non-interacting case.

\item An attraction speeds up the reaction generally.
Outside the reaction volume, it increases the encounter rate $k_\mathrm{e}$; here, the sign of $U(r) - U(r \to \infty)$ matters, which points at an energetic origin.
For the formation rate $k_\mathrm{f}$, the force $-U'(r)$ inside the reaction volume and the value $p(R)$ on the boundary enter.

\item For mixed situations as for the LJ potential, both contributions, $k_\mathrm{e}$ and $k_\mathrm{f}$, are non-monotonic in the position of the reaction boundary (\cref{fig:ratesEnhancement}) and can lead to non-trivial dependencies of the total rate $k$ on the model parameters $\lambda$ and $R$ (\cref{fig:partialRates}).
\end{enumerate*}

Concluding, we have established a microscopic simulation model that extends Doi's volume reaction model to interacting molecules.
This model is at the core of iPRD simulations, which permit treatment of spatially resolved reaction processes in cells and nanotechnology at different levels of coarse graining.
The obtained relation between $k$ and the parameters $\lambda, R$ facilitates the development of quantitative iPRD models based on experimental values of the macroscopic rate $k$.
The interaction potential $U(r)$, can either be chosen \emph{ad hoc} based on physical insight or determined as the potential of mean force in atomistic simulations \cite{buch2011optimized,Xu:2019,wu2016multiensemble}.
The freedom to choose an interaction potential within the reaction volume offers the opportunity to implement coarse-grained simulations that switch between representations of bound complexes using either explicit potential wells and barriers or stochastic reactions.
The present study focuses on the dilute limit, which serves as a well-defined starting point for
the investigation of concentration and crowding effects on the reaction rate and the distribution of molecules.

\begin{acknowledgments}
This research has been funded by Deutsche Forschungsgemeinschaft  through
grants SFB 1114 (project C03) and SFB 958 (project A04) and under
Germany's Excellence Strategy -- MATH+: The Berlin Mathematics Research
Center (EXC-2046/1) -- project ID: 390685689 (subproject AA1-6).
Further funding by the Einstein Foundation Berlin (ECMath, project CH17)
and by the European Research Council (ERC CoG 772230 ``ScaleCell'') is
gratefully acknowledged.
\end{acknowledgments}

\bibliography{literature/used_citations}

\end{document}